\documentclass[letterpaper, 10 pt, conference]{ieeeconf}  % Comment this line out
\IEEEoverridecommandlockouts                              % This command is only
\usepackage{graphicx} % for pdf, bitmapped graphics files
\usepackage{epsfig} % for postscript graphics files
\usepackage{times} % assumes new font selection scheme installed
\usepackage{amsmath} % assumes amsmath package installed
\usepackage{amssymb}  % assumes amsmath package installed
\usepackage{blindtext}
\usepackage{algpseudocode}
\usepackage{upgreek}
\usepackage{physics}
\usepackage{hyperref}
\usepackage{xcolor}
\usepackage{cancel}
\usepackage{cleveref}
\usepackage[normalem]{ulem}
\usepackage{algorithm}
\newcommand{\edits}[1]{\textcolor{black}{#1}}

\newif\ifhideedits
\hideeditsfalse

\DeclareMathOperator*{\E}{\mathbb{E}}

\newcommand{\vect}[1]{\mathrm{vec}\left(#1\right)}

\title{\LARGE \bf
	Control-Oriented Identification for the Linear Quadratic Regulator: Technical Report}

\author{Sean Anderson, João P. Hespanha% <-this % stops a space
	\thanks{This material is based upon work supported by the National Science Foundation Graduate Research Fellowship under Grant No. 2139319 and by the U.S. Office of Naval Research MURI grant No. N00014-23-1-2708. Any opinion, findings, and conclusions or recommendations expressed in this material are those of the authors(s) and do not necessarily reflect the views of the National Science Foundation or U.S. Office of Naval Research.}% <-this % stops a space
	\thanks{The authors are with Department of Electrical and Computer Engineering, University of California Santa Barbara, Santa Barbara, CA, 93106 USA
		{\tt\small seananderson@ucsb.edu, hespanha@ece.ucsb.edu}}%
}

\begin{document}

	\maketitle
	\thispagestyle{empty}
	\pagestyle{empty}
	
	%%%%%%%%%%%%%%%%%%%%%%%%%%%%%%%%%%%%%%%%%%%%%%%%%%%%%%%%%%%%%%%%%%%%%%%%%%%%%%%%
	\begin{abstract}
		Data-driven control benefits from rich datasets, but constructing such datasets becomes challenging when gathering data is limited. We consider an offline experiment design approach to gathering data where we design a control input to collect data that will most improve the performance of a feedback controller. We show how such a control-oriented approach can be used in a setting with linear dynamics and quadratic objective and, through design of a gradient estimator, solve the problem via stochastic gradient descent. We show our formulation numerically outperforms an A- and L-optimal experiment design approach as well as a robust dual control approach.
	\end{abstract}

	%%%%%%%%%%%%%%%%%%%%%%%%%%%%%%%%%%%%%%%%%%%%%%%%%%%%%%%%%%%%%%%%%%%%%%%%%%%%%%%%
	\section{Introduction}

	Model-based control methods benefit from accurate models of the controlled system. Consider a setting in which there is uncertainty in the model parameters and there is an opportunity to collect experimental data to learn more about the system. This motivates the following control-oriented experiment design problem: select a control input for a data-collection experiment so that the feedback controller designed using the data acquired will lead to \edits{improving the control performance as much as possible}. 
	
	This paper includes two key contributions: \edits{First, we propose an experiment design formulation that explicitly optimizes the post-experiment closed-loop control performance. Notably, this formulation sidesteps the classical exploration-exploitation tradeoff through a unique optimization and achieves ``optimal" exploration by construction. Second, we derive a gradient estimator to solve the resulting nonconvex optimization through stochastic gradient descent.} In the setting with linear dynamics and quadratic objective function, we observe that our method generally leads to closed-loop controllers that exhibit higher performance than what would be achieved by 1) classical forms of experiment design \edits{such as A- \cite{chernoff_locally_1953} and L-optimal design and 2) a robust dual control method that minimizes the worst-case system cost.}
	
%	This paper includes two key contributions: First, we show how such a control-oriented approach to experiment design can be carried out for the control of a linear system with unknown matrix dynamics and a quadratic objective function. While this problem does not have a closed-form solution, we show that it can be efficiently solved by stochastic gradient descent. The second contribution lies in the observation that our formulation of a control-oriented approach to experiment design generally leads to closed-loop controllers that exhibit higher performance than what would be achieved by 1) classical forms of experiment design \edits{such as A- \cite{chernoff_locally_1953} and L-optimal design}, rather than focusing directly on improving closed-loop performance; \edits{and 2), a robust dual control method that minimizes the worst-case system cost.}
	
	In Section \ref{sec:oed_general} we present a general formulation for experiment design that aims to minimize the expected post-experiment control performance by taking into account 1) $\emph{a-priori}$ parameter uncertainty in discrete-time dynamics and 2) process disturbances that occur during the experiment. Our approach is general in terms of the control design procedure used to generate the controller from the experimental data collected. However, in this paper we focus our attention on controllers generated through certainty equivalence, which in this context means constructing an \emph{a-posteriori} estimate for the process and designing a controller for this estimate. 
	
	We present the solution method to the experiment design problem in Section \ref{sec:sgd}. We use a first-order approach by designing a pathwise gradient estimator for the purposes of stochastic gradient descent. For the remainder of the paper, we focus on the linear quadratic regulator (LQR) setting presented in Section \ref{sec:oed_lqr}. We address the system identification step and how to handle exploding trajectories during (simulated) experiments. For the data-driven controller, we use an LQR controller with certainty equivalence in the parameters \cite{simon_dynamic_1956}. The solutions available in LQR are amenable to fast computations, which is conducive to the gradient estimator presented in \ref{sec:sgd}. In Section \ref{sec:num_results}, we compare our method against A- and L-optimal experiment design in a car string setting and show how our approach scales numerically. We also consider how our approach can be compared against a recent robust dual control formulation.
	
	\textit{Related work:} The issue of how to gather data through well-planned experiments has traditionally been addressed through the framework of optimal experiment design. Modern optimal experiment design is often attributed to Gustav Elfving, who designed experiments to minimize measures of parameter error covariance \cite{elfving_optimum_1952}. Later on, researchers worked on aligning experiment design with particular criteria, including control objectives \cite{lindqvist_identification_2001, gevers_identification_2005}. Recent work in this area includes \cite{anderson_cdc2023}, which proposes a stochastic gradient descent approach to designing experiments that minimizes a post-experiment optimal control objective. Work in experiment design for control in the statistical learning community includes \cite{lee2023fundamental, dean_sample_2020}, which emphasize theoretical aspects of learning linear systems. As an alternative paradigm, online learning \cite{simchowitz_naive_2020} or adaptive control \cite{astrom_adaptive_2008} allow for improvement during the experiment trial. 
	
%	\edits{\sout{As we focus on the linear quadratic regulator setting, the most relevant papers are from \cite{lee2023fundamental}, in which the authors derive fundamental limits for learning the linear quadratic regulator via offline experiments, and \cite{wagenmaker_task-optimal_2021}, in which the authors propose an experiment design method for learning linear systems for particular tasks. The former focuses on the statistical learning guarantees associated with the LQR problem but do not broach how to construct an optimal experiment. In the latter, the authors propose a weighted-trace optimal experiment design that bears similarity to $L$-optimal experiment design for control-oriented identification as noted in \cite{gevers_identification_2005}; the authors consider convergence of sequential experiment designs to  the underlying system.}}
	
	\edits{In the control community, recent work in the linear quadratic setting includes \cite{umenberger2019, ferizbegovic2020} in which the authors propose a robust dual control approach that minimizes the control cost associated with a worst-case system.  The authors in \cite{venkatasubramanian2020} consider a robust gain-scheduling approach while \cite{rallo2018} proposes a robust experiment design method for virtual reference feedback tuning.}
	
	Gradient estimation has seen particular attention in the machine learning community \cite{mohamed_monte_2020}, and two dominant methods are via the pathwise gradient and the score function (or log score) method \cite{peters_reinforcement_2008}. Score function estimators benefit from only taking the gradient of the density function, but the structure of our problem is not amenable: as such, we focus on the pathwise gradient estimate.
	
	\section{Experiment Design for Data-Driven Control}   \label{sec:oed_general}
		
		We consider a discrete-time system with dynamics of the form
		\begin{align} \label{eq:dynamics}
			x_{t+1} = f(x_t, u_t, w_t; \theta),
		\end{align}
		with state $x_t \in \mathbb{R}^{n_x}$, control input $u_t \in \mathbb{R}^{n_u}$, and an unmeasured stochastic disturbance $w_t \in \mathbb{R}^{n_w}$ independent and identically distributed across time. The dynamics depend on parameters $\theta$ that are unknown, but for which we have an \emph{a-priori} distribution. \edits{As such, we treat $\theta$ as a random variable in the sense that we do not know its value during an experiment realization.}
		
		An experiment will be performed to provide additional information about the parameter $\theta$. State and input measurements are collected throughout the experiment providing a sequence of $M$ triples $\mathcal{D}:=\{(x_i^+, x_i, u_i): i=1,...,M\}$ that satisfy the basic model of the dynamical system:
		\begin{align} \label{eq:dataset}
			x_i^+ = f(x_i, u_i, w_i; \theta), \quad\forall i\in\{1,\dots,M\}
		\end{align}
		with the $w_i$ independent across indices $i$ and with the same distribution as the disturbance. If the experiment consists of a single run of \eqref{eq:dynamics} over a time horizon $t=1$ through $t=T$, then the index $i$ is simply time and $M=T$. However, in general, ``an experiment" may include multiple runs of \eqref{eq:dynamics} over different time horizons, in which case \eqref{eq:dataset} include all the data collected. To simplify notation we collect all the columns vectors $x_i^+,x_i \in \mathbb{R}^{n_x}, u_i \in \mathbb{R}^{n_u}$ into matrices with $M$ columns that we denote by $X^+$, $X \in \mathbb{R}^{n_x \times M}, U \in \mathbb{R}^{n_u \times M}$, respectively.
		
		Our goal is to design a controller $\pi$ that optimizes a given cost function $J(\pi; \theta)$ that depends both on the controller and \edits{$\theta$, which at the time of control synthesis we only have an a-posteriori distribution for, given $\mathcal{D}$}. We also take as given a control design procedure that maps the experiment design data $\mathcal{D}$ to a specific controller $\pi$, with the goal of minimizing the cost $J(\pi;\theta)$. \edits{In this work we consider a controller $\pi = K(\mathcal{D})$ that minimizes $J(\pi;\hat{\theta})$ where $\hat{\theta} := \E_\theta [\theta \,|\, \mathcal{D}]$, but our approach allows for general control design procedures $K:\mathcal{D} \mapsto \pi$.}
	%	\edits{\sout{, as in [FIX]}}:
	%	\begin{align} \label{eq:gen_controller}
	%		\edits{[REMOVE]} K(\mathcal{D}) = \argmin_\pi \E_{\edits{\theta}}[J(\pi;\theta)\,|\, \mathcal{D}]
	%	\end{align}
	%	\edits{\sout{where $\E_{\theta}[J(\pi;\theta)\,|\, \mathcal{D}]:= \int J(\pi; \theta) p( \theta \,|\, \mathcal{D}) d\theta$.}} 
%		However, \edits{\sout{and because this optimization is often intractable,}} our presentation considers a general control design method $K:\mathcal{D} \mapsto \pi$, which may or may not be optimal.
		
		The experiment design problem arises from the observation that the data $\mathcal{D}$ collected depends on the \edits{realization of the parameter $\theta$,} control inputs $U$ used during the experiment, as well as on the realizations of the random disturbances $w_t$ \edits{such that the state trajectory $X$ is a random variable.} We use the notation $\mathcal{D}_{U,\edits{X }}$ to express the dependence of the dataset on these variables. The optimal experiment design problem can then be formulated as 
		\begin{align}\label{eq:exp_des}
			\min_{U\in\mathcal{U}} \E_{\edits{X,\theta}}\big[ J\big(\pi;\theta\big) \big], \quad \pi:= K\big(\mathcal{D}_{U,\edits{X }}\big),
		\end{align}
		where \edits{$\E_{X,\theta}\big[ J\big(\pi;\theta\big) \big]:= \int J\big(\pi;\theta\big) p(X, \theta; U) dX d\theta$} refers to an integration over (i) the \emph{a-priori} distribution $p(\theta)$ of the parameter $\theta$, and (ii) the realization of the \edits{
		state trajectory} during the experiment \edits{of length $T$}. The minimization is
		performed over a set of admissible controls that we denote generically
		by $\mathcal{U}$.
		
	%	\edits{\sout{For example, an A-optimal experiment design
	%	essentially tries to minimize \edits{FIX expectation}}}
	%	\begin{align}\label{eq:A_exp_des}
	%		\min_{U\in\mathcal{U}}\E_{\edits{??}}\Big[ \big\|\hat\theta\big(\mathcal{D}_{U,\edits{X }}\big) -\theta\big\|^2 \Big], \quad
	%		\hat\theta\big(\mathcal{D}_{U,\edits{X }}\big) := \E_\theta \big[\theta\,|\,\mathcal{D}_{U,\edits{X }}\big].
	%	\end{align}
	%	\edits{\sout{The key distinction is that \eqref{eq:A_exp_des} ignores the
	%	impact of uncertainty on the control objective $J(\pi;\theta)$ and therefore
	%	does not take advantage of the fact that reducing uncertainty on some
	%	parameters may be much more important than on others, \emph{for our ultimate control objective of minimizing $J(\pi; \theta)$.}}}
		\section{\edits{Experiment design via gradient descent}} \label{sec:sgd}

	In order to solve the experiment design optimization \eqref{eq:exp_des}, we take a gradient-descent approach:
	\begin{align} \label{eq:grad_desc}
		U_{i+1} = \text{Proj}_\mathcal{U}(U_{i} - \eta_i \hat{\nabla}_U), 
	\end{align} 
	where $\text{Proj}_\mathcal{U}(U)$ projects $U$ onto the set of admissible inputs $\mathcal{U}$, $\eta_i$ is step size, and we estimate the true gradient $\nabla_U$ with a pathwise gradient estimator to produce $\hat{\nabla}_U$.
	
%	 given by differentiating the experiment criteria in \eqref{eq:exp_des}:
%	\begin{align} 
%		\begin{split} \label{eq:gen_integral}
%			%		\nabla_U \E_{X, \theta}\Big[ &J\Big(K\big(\mathcal{D}\big); \theta\Big) \Big] = \\
%			&   \nabla_U \int J\big(K\big(\mathcal{D}\big);\theta\big) p(X \,|\, U , x_0,  \theta) p(\theta)  d{\theta} dX,
%		\end{split}
%	\end{align}
%	where $x_0$ is the initial state for the experiment. The gradient of the integral is analytically intractable and of dimension $n_x(n_x+n_u)+n_wT$, motivating the use of a Monte Carlo gradient estimate over methods such as quadrature. Gradient estimation typically involves exchange of the integral and gradient such that, if justified,
%	\begin{align}
%		\begin{split}
%			\nabla_U \int & J\big(K(\mathcal{D}_{U,X});\theta\big) p(X \,|\, U, \theta) p(\theta) dX d\theta \\
%			&= \int \nabla _U\bigg(J\big(K(\mathcal{D}_{U,X});\theta\big) p(X \,|\, U, \theta) p(\theta) \bigg) dX d\theta.
%		\end{split}
%	\end{align}
	
	We recall that for a general function $F(y)$ that is differentiable with respect to a random variable $y$ with probability density function $p(y;U)$ that depends on a parameter $U$, the Monte Carlo pathwise gradient estimator of 
	\begin{align} \label{eq:pathwise_init}
		\nabla_U \E_y[F(y)] := \nabla_U \int F(y) p(y ; U)  dy
	\end{align}
	is defined by
	\begin{align} \label{eq:pathwise_mc}
		\hat{\nabla}_U := \frac{1}{L} \sum_{l=1}^L \nabla_U F(g(\epsilon^{(l)}; U)), ~~\epsilon^{(l)} \sim p(\epsilon),
	\end{align}
	where $L$ is the number of Monte Carlo samples, and $g$ is a differentiable sampling path such that $y$ is distributed the same as $z := g(\epsilon; U)$, where  $\epsilon$ is a random variable with continuous distribution $p(\epsilon)$ \cite{mohamed_monte_2020}. Pathwise gradient estimators are unbiased, typically low variance, and computationally efficient \cite{mohamed_monte_2020}. The variance has been shown to be bounded by the square of the Lipschitz constant of $F$ \cite{fan_fast_2017}. 
	
	\textit{Assumption (Regularity) 1}: Assume that the controller $\pi$ is parameterized by S scalar parameters $(\pi_1,\pi_2,…,\pi_S)$; $J$ is differentiable with respect to $\pi_s$ such that infinitesimal perturbations in $\pi_s$ lead to infinitesimal perturbations in $J$, where $\pi_s$ comes from the control design procedure $K(\mathcal{D})$; the procedure $K(\mathcal{D})$ is differentiable with respect to $X$ and $U$ such that infinitesimal perturbations in the data lead to infinitesimal perturbations in the controller; and, similarly, $f(\cdot)$ in \eqref{eq:dynamics} is differentiable with respect to $x_t$ and $u_t$. 
	
	\textit{Theorem 1}: Assume that the process noise at each time step is distributed according to a continuous distribution $p(w_t)$ such that $p(W):=\Pi_{t=1}^{T-1}p(w_{t-1})$ and Assumption 1 holds. Then, the $ij$th element of the pathwise gradient estimator of $\E_{X,\theta}[J(\pi;\theta)]$ in \eqref{eq:exp_des} is given by
	\begin{align} \label{eq:pw_mc}
%		\hat \nabla_U = \frac{1}{L} \sum_{l=1}^L \nabla_U J\Big(K\big(\mathcal{D}_{U,g(W^{(l)},\theta^{(l)}; U)}\big); \theta^{(l)}\Big),
	\hat \nabla_{U_{ij}} = \frac{1}{L} \sum_{l=1}^L \sum_{s,m,n} \pdv{J}{\pi_{s}}\bigg(\pdv{K_{s}}{X_{mn}}\pdv{g_{mn}}{U_{ij}} + \pdv{K_{s}}{U_{ij}}\bigg),
	\end{align}
	with the understanding that the gradient is evaluated at $U$, and the inner sum is over all $S$ parameters and the elements of $X$. The sampling path is given by:
	\begin{subequations}
		\begin{align}
			g(W, \theta;U) = \begin{bmatrix}
				x_0 \\
				g_0(x_0, u_{0}, w_{0}; \theta) \\
				\vdots \\
				g_{T-2}(x_0, u_{0:T-2}, w_{0:T-2}; \theta)\\
			\end{bmatrix},
		\end{align}
		such that we sample $W^{(l)}$ from $p(W)$ and $\theta^{(l)}$ from $p(\theta)$, and $u_{0:k}$ are the first $k+1$ columns of $U$, $g_{k}(x_0, u_{0:k}, w_{0:k}, \theta)$$:=f(g_{k-1}(x_0,u_{0:k-1},w_{0:k-1}), u_k, w_k; \theta)$ $\forall k \in \{1,...,T-1\}$, and $g_0(x_0, u_0, w_0; \theta) = f(x_0, u_0, w_0, \theta)$.
	\end{subequations}
	\newline
	\begin{proof}
		\begin{subequations} 
			Consider the integral in the experiment design criteria in \eqref{eq:exp_des}:
			\begin{align} 
				\int J\big(K(\mathcal{D}_{U,X});\theta\big) p(X , \theta; U) dX d\theta. \label{eq:exp_int}
			\end{align}
		 By Bayes' rule for probability density functions $p(X, \theta; U) = p(X \,|\, \theta; U) p(\theta; U)$ and the joint distribution of the states can be recursively expanded as 
			\begin{align}
				p(X \,|\, \theta; U) &= \Pi_{t=1}^{T-1} p(x_t \,|\, x_0, ..., x_{t-1}, \theta; U),
			\end{align}
		where $x_0$ is known. From \eqref{eq:dynamics}, we have Markovian dynamics such that $p(x_t \,|\, x_0,..., x_{t-1}, \theta; U) = p(x_t \,|\, x_{t-1}, \theta; u_{t-1})$ and 
			\begin{align}
				&p(X \,|\, U, x_0, \theta) = \Pi_{t=1}^{T-1} p(x_t \,|\, x_{t-1}, \theta; u_{t-1}). \\
				\intertext{We can further decompose this by integrating over the process noise in \eqref{eq:dynamics}:}
				&=\Pi_{t=1}^{T-1} \int p(x_t \,|\, x_{t-1}, w_{t-1}, \theta; u_{t-1}) p(w_{t-1}) dw_{t-1} .
			\end{align}
			Since $p(x_t \,|\, x_{t-1}, w_{t-1}, \theta; u_{t-1})$ occurs with probability one when the state at time $t$ equals $x_t$, we express this using a delta function:
			\begin{align}
				&=\Pi_{t=1}^{T-1} \int \delta(x_t - f(x_{t-1}, u_{t-1}, w_{t-1})) p(w_{t-1}) dw_{t-1}.
			\end{align}
			Substituting into \eqref{eq:exp_int} yields
			\begin{align}
				\begin{split}
					\int &J\big(K(\mathcal{D}_{U,X}); \theta\big) \Pi_{t=1}^{T-1} \int \delta\big(x_t - f(x_{t-1}, u_{t-1}, w_{t-1})\big) \\
					&p(w_{t-1}) dw_{t-1} dx_{t-1} p(\theta) d\theta.
				\end{split} 
			\end{align}
		Integrating with respect to any $x_t$ leads to
			\begin{align}
				\begin{split}
					&\int J\big(K(\mathcal{D}_{U,X}); \theta\big)\delta\big(x_t - f(x_{t-1}, u_{t-1}, w_{t-1})\big) dx_t = \\ 
					&J\big(K(\mathcal{D}_{U, [x_0,...x_t=g_{t-1}(x_0, u_{0:t-1}, w_{0:t-1}; \theta),...]^T}; \theta\big).
				\end{split}
			\end{align}
			If we integrate out $x_t$ for all $t$ and define $g(W, \theta; U) := [x_0, g_0, ..., g_{T-1}]^T$, then \eqref{eq:exp_int} equals
			\begin{align} \label{eq:ind_int}
				\int J\big(K(\mathcal{D}_{U, g(W,\theta;U)}); \theta\big) \Pi_{t=1}^{T-1} p(w_{t-1}) dw_{t-1} p(\theta)  d\theta,
			\end{align}
			such that the probability density functions are independent of $U$. Thus, under the change of variable $g$, $\mathbb{E}_{X,\theta}\big[J\big(K(\mathcal{D}_{U,X}; \theta)\big)\big] = \mathbb{E}_{W,\theta} \big[J\big(K(\mathcal{D}_{U, g(W,\theta;U)})\big)\big]$. Now, differentiating \eqref{eq:ind_int} can be achieved under the differentiability assumptions on $J$ and $K$, and $g$ inherits differentiability from $f$. Differentiating $J$ with respect to the $ij$th element of $U$ at the current input $U$ and $X=g(W,\theta;U)$ yields
			\begin{align} 
				\nabla_{U_{ij}} J = \sum_{s,m,n} \pdv{J}{\pi_{s}}\bigg(\pdv{K_{s}}{X_{mn}}\pdv{g_{mn}}{U_{ij}} + \pdv{K_{s}}{U_{ij}}\bigg).
			\end{align}
		We then take a Monte Carlo sample average to obtain \eqref{eq:pw_mc}.
			\end{subequations}
	\end{proof}
	
	\subsection{Algorithm for Experiment Design Problem}
	In the pathwise gradient estimator \eqref{eq:pw_mc} for each sample, $l$, we obtain a single experiment trajectory under sampled noise $W$ for a sampled system $\theta$ under the candidate input $U$. For this realization, we compute an \emph{a-posteriori} system estimate, $\hat{\theta}$, and compute the control, $\pi$. \edits{The step size $\eta_i$ decays exponentially, and Algorithm 1 terminates when the moving average of the norm of the gradient is sufficiently small or hits the max allowable iterations.}
	
	\edits{The main hurdles in this general setting are obtaining a computationally efficient form of 1) $K(\mathcal{D})$ since this requires estimating the system and computing a controller with respect to the system estimate and 2) $J(\cdot)$ due to computing the value function and $K(\mathcal{D})$ for each sample.}
	
	\begin{algorithm}
		\caption{\strut Control-Oriented Experiment Design}\label{alg:exp_des_algo}
		\label{alg:exp_design}
		\hspace*{\algorithmicindent} \textbf{Input} $p(\theta) \text{ (prior on } \theta\text{)}, ~U_0 ~ (\text{initialization}),$ \par
		\hspace*{\algorithmicindent} $L ~\text{(batch size)}, ~\mathcal{U} ~\text{(feasible set)}, ~p(W) ~\text{(noise dist.)}$ \par
		\hspace*{\algorithmicindent} \textbf{Output} $U^*$
		\begin{algorithmic}
%			\Function{\edits{$K$}}{$\mathcal{D}$} 
%			\State $\hat{\theta} \gets$ \text{estimate system given}  $\mathcal{D}$ 
%			\State $K \gets$  \text{Solve control problem given $\hat{\theta}$}
%			\EndFunction
			
%			\Function{\edits{$\nabla J$}}{$U, \theta, W$} 
%			\EndFunction
			
			\While{not converged}
			\For{i=1 to L}
			\State $\theta^{(i)} \sim p(\theta), W^{(i)}  \sim p(W)$ 
			\State $X \gets g(W^{(i)}, \theta^{(i)}; U_j)$
			\State $\mathcal{D} \gets X, U_j$
			\State $\pi \gets \Call{$K$}{\mathcal{D}}$
			\State $\nabla_U J_i  \gets$  \text{Compute gradient of $J(\pi; \theta^{(i)} )$}
			\EndFor
			\State $U_{j+1} \gets \text{Proj}_\mathcal{U}\big(U_j - \eta_j \frac{1}{L} \sum_{l=1}^L \nabla_U J_l\big)$
			\EndWhile
		\end{algorithmic}
	\end{algorithm}
	
	\section{Experiment Design for the Linear Quadratic Regulator} \label{sec:oed_lqr}
	We now specialize the general setup described above to the finite-horizon linear quadratic regulator setup. Specifically, we consider the process
	\begin{align}
		x_{t+1} = Ax_t + Bu_t + w_t, \label{eq:linear_dynamics}
	\end{align}
	such that $\theta$ contains the elements of $A$ and $B$, and $w_t$ is Gaussian noise identically distributed across time with zero mean and covariance $\Sigma_w$.
	
	We consider a quadratic optimization criterion of the form:
	\begin{align}  \label{eq:lqr}
		J(\pi; \theta) := \mathbb{E}_{\edits{W}}\bigg[x_N^TQ_N x_N + \sum_{t=0}^{N-1}x_t^TQ x_t + u_t^T R u_t \,|\, \theta \bigg],
	\end{align}
	where the expectation refers to an integration over the disturbances\edits{, $W:=[w_0,...,w_{N-1}]$,} encountered by the controller $\pi$; $Q_N, Q$ are positive semidefinite matrices; and $R$ a positive definite matrix.
	
	We consider a common option for control design generally known as \emph{certainty equivalence (CE)}: certainty equivalence design $K_{CE}(\mathcal{D})$ computes the \emph{a-posteriori} expected value of the unknown
	parameters $\hat\theta:= \E_\theta[\theta \,|\, \mathcal{D}]$ and computes the linear
	optimal controller $u_t = K_t x_t$ that minimizes \eqref{eq:lqr},
	\emph{assuming that the estimate $\hat\theta$ is correct}.
	
	\subsubsection{System identification}
	In order to generate the \emph{a-posteriori} estimate of the system $\hat \theta$ for $K_{CE}(\mathcal{D})$, we employ weighted Bayesian estimation on a dataset $\mathcal{D}$, which in our case will be the dataset generated under the experiment decision variable $U$. For identification, we express \eqref{eq:linear_dynamics} as:
	\begin{align}
		X^+ = \Theta Z + W
	\end{align}
	with $Z = [X; U] \in \mathbb{R}^{(n_x+n_u) \times M}$, and $\Theta:=[A,B] \in \mathbb{R}^{n_x \times (n_x + n_u)}$. For ease of notation, we use $\theta \in \mathbb{R}^{n_x(n_x+n_u)}$ to denote the vectorized version of $\Theta$ via stacking its columns.
	
	We consider a Gaussian prior on the parameters with mean $\Theta_0 \in \mathbb{R}^{n_x \times (n_x+n_u)}$ and covariance of the $(i,j)$th element with the $(k,l)$th element of $\Theta$ given by $\E_\Theta[(\hat{\Theta} - \Theta)_{ij}(\hat{\Theta}-\Theta)_{kl}] = (\Sigma_w)_{ki} (\Lambda_0^{-1})_{jl} $, where $\Sigma_w$ is the known noise covariance and $\Lambda_0^{-1} \in \mathbb{R}^{(n_x+n_u) \times (n_x+n_u)}$ is a prior on the parameter covariance. The weighted Bayesian estimator for $\Theta$ is
	\begin{subequations} \label{eq:estimator}
		%	\vspace{-0.2em}
		\begin{align}
			\hat{\Theta} = (\Theta_0 \Lambda_0+ X^+SZ^T) \Lambda_n^{-1}, 
		\end{align}
		and the error covariance of the estimate $\hat{\Theta}$ is
		\begin{align} \label{eq:error_cov}
			\E_\Theta\big[(\hat{\Theta} &- \Theta)_{ij}(\hat{\Theta}-\Theta)_{kl}\big] = (\Sigma_w)_{ki} (\Lambda_n^{-1})_{jl},
		\end{align}
		where $\Lambda_n := \Lambda_0 + ZSZ^T$, and $S \in \mathbb{R}^{M \times M}$ is the weight matrix. \edits{While a different prior and estimator could be used, this closed-form solution allows for efficient computations for our proposed experiment design. For derivation see e.g \cite{rossi2006}}.
	\end{subequations}
	
	The weight matrix $S$ improves the numerics of the regression problem, particularly since simulating unstable systems can lead to exponential growth in the state that, due to large numbers, lead to deleterious performance in the inversion of $\Lambda_n$. In particular, let 
	\begin{align}
		S(X) := \text{diag}([s(x_0),...,s(x_N)]) 
	\end{align}
	where $s(x)\in [0,1]$ ensures that the weight matrix assigns zero weight to points on trajectories that are numerically too large. For this work, we choose $s(x) := \arctan(\norm{x_t-\alpha_1}\alpha_2)/\pi + 0.5$, and $\alpha_1, \alpha_2$ are design parameters.
	
	\subsubsection{Certainty equivalent control}
	Given an estimate of the parameters $\theta$ from \eqref{eq:estimator} with means $\hat{A}$ and $\hat{B}$, respectively, we construct our controller $K_{CE}(\mathcal{D})$ by recursively solving the Riccati difference equations given by
	\begin{subequations}  \label{eq:ricc_diff}
		\begin{align}
			K_t &= -(R+\hat{B}^TP_{t+1}\hat{B})^{-1}\hat{B}^TP_{t+1}\hat{A}, \\
			P_t &= Q + K_t^TRK_t + (\hat{A}+\hat{B}K_t)^TP_{t+1}(\hat{A}+\hat{B}K_t), 
		\end{align}
	\end{subequations}
	with $P_N = Q_N$; $Q,Q_N$ are positive semidefinite matrices and $R$ a positive definite matrix. 
	
	\textit{Corollary \edits{1}}:  \cite{bertsekas_dynamic_2012} For a sequence of linear feedback gains, $\pi:=\{K_0,...,K_{N-1}\}$ from $K_{CE}(\mathcal{D})$, we can express the finite-horizon LQR cost \eqref{eq:lqr} for the system in \eqref{eq:linear_dynamics} parameterized by $\theta$ as
	\begin{subequations}
		\begin{align} \label{eq:closed_form_cost}
			J(\pi; \theta) &= x_0^TP_0x_0 + \sum_{t=0}^{N-1}\trace(P_{t+1}\Sigma_w),
		\end{align} where
		\begin{align}
			P_t &= Q + K_t^TRK_t + (A+BK_t)^TP_{t+1}(A+BK_t),
		\end{align}
		with boundary condition $P_N = Q_N$.
	\end{subequations}
	
\textit{\edits{Corollary 2}}: For the LQR experiment design pathwise gradient estimate, the sampling path $g(W, \theta; U)$ is given by
\begin{subequations} \label{eq:lqr_estimator}
\begin{align}
	g(W, \theta;U) = \begin{bmatrix}
		x_0 \\
		A x_0 + B u_0 + w_0 \\
		\vdots \\
		A^{T-1}x_0 + \sum_{l=0}^{T-2}A^{T-2-l}(Bu_l + w_l)\\
	\end{bmatrix},
\end{align}
and in this problem the controller can be parameterized by the controller gains or the certainty equivalent estimate from which the gains are constructed. 
%and the $(i,j)th$ element of the $l$th sample of $\hat{\nabla}$ is given by
%\begin{align}
%	\hat{\nabla}_{U_{ij}}J = \sum_{t,q,r,m,n} \pdv{J}{K_{tqr}} \pdv{K_{tqr}}{\hat{\theta}_{mn}}  \pdv{ \hat{\theta}_{mn}}{U_{ij}}.
%\end{align}
%where $K_t$ is given by \eqref{eq:ricc_diff} and $\hat{\theta}$ is given by \eqref{eq:estimator}.

\end{subequations}
\begin{proof}
	See Appendix \ref{sec:linear_change_of_var} for the derivation of $g$ and Appendix \ref{sec:diff_app} for the gradient expression.
\end{proof}

In practice, the gradient can be computed efficiently using automatic differentiation.

%	\textit{Lemma 2}: The change of variable \eqref{eq:g_linear} and LQR experiment criteria \eqref{eq:lqr}, with $K_{CE}(\mathcal{D})$, yields an estimator of the form \eqref{eq:pathwise_mc}:
%	\begin{align}
%		\hat \nabla_U = \frac{1}{L} \sum_{l=1}^L \nabla_U J\Big(K\big(\mathcal{D}\rvert_{X=G(W^{(l)}; U, x_0, \theta^{(l)})}\big); \theta^{(l)}\Big),
%	\end{align}
%	with $W^{(l)} \sim p(W).$

%\begin{proof}
%	See Appendix \ref{sec:diff_app}.
%\end{proof}

	\section{Numerical Experiments} \label{sec:num_results}
	
	\subsection{Car String}
	
%	\begin{figure}[th]
%		\begin{center}
%			\includegraphics[width=8.4cm]{media/car_string.png}    % The printed column width is 8.4 cm.
%			\caption{We focus on a scenario in which cars need to regulate to a specified gap $\bar{L}$ at a desired reference velocity $v$. Deviations from the desired position are $\Delta w^{(n)}$ and deviations from the reference velocity are $\Delta v^{(n)}$.}
%			\label{fig:car_string}
%		\end{center}
%	\end{figure}
	
	We consider the problem of maintaining a fixed distance, $\bar{L}$, between $n$ cars at a desired velocity $v$. We adapt the continuous-time dynamics for relative position as given in \cite{levine_optimal_1966} to discrete-time dynamics with sampling time $T_s$:
	\begin{align}
		\Delta v_{t+1}^{(n)} &= \bigg(-\frac{\alpha^{(n)} T_s}{m^{(n)}}  + 1\bigg)\Delta v_t^{(n)} + \frac{T_s}{m^{(n)} } \Delta u_t^{(n)}, \\
		\Delta w_{t+1}^{(n)} &= T_s(\Delta v_t^{(n)}- \Delta v_t^{(n+1)}) + \Delta w_t^{(n)},
	\end{align}
	where $\Delta v^{(n)}$ is the deviation from the reference velocity at car $n$ and $\Delta w^{(n)}$ is the deviation of the gap between cars $n+1$ and $n$ from the desired gap $L$. $\Delta u$ is a change in force input for each car. This leads to an $n$ car state-vector $x_{t+1}:=[\Delta v_{t+1}^{(1)}, \Delta w_{t+1}^{(1)}, \Delta v_{t+1}^{(2)},...,\Delta v_{t+1}^{(n)}]^T$.
	While there is a specific structure to the resulting $(A,B)$ matrices, we assume we do not know the structure and estimate all $(2n-1)(3n-1)$ entries \edits{as our method does not require \emph{a-priori} knowledge of structure}. We specify the noise covariance in the dynamics \eqref{eq:linear_dynamics} as $\Sigma_w = 1\mathrm{e}{-2} \times I_5$. 
	The prior on the parameters \eqref{eq:estimator} is $\Theta_0=[A,B]$ with $m^{(1)}=m^{(2)}=m^{(3)}=1$, $\alpha^{(1)}=\alpha^{(2)}=\alpha^{(3)}=1$, $T_s=0.1$; \edits{$\Lambda_0^{-1}=\text{diag}([0.1, 0.01, 0.05, 0.1, 0.01, 0.05, 0.1, 0.05])$, motivated by having high uncertainty in the velocity evolution and the influence of the input. The full expressions for $A$ and $B$ in the problem are shown in Appendix \ref{sec:car_string_app} of \cite{anderson_tech24}. Horizon $N=30$}.
	
	\subsubsection{Experiment Design Setup}
	In the results that follow, we use an experiment horizon of $T=20$ time steps, and batch size $L=1000$ in Algorithm 1. As in \cite{levine_optimal_1966}, for the criteria in \eqref{eq:lqr} $Q$ includes penalties of magnitude $10$ on the positions $\Delta w$ and zero on the velocity $\Delta v$. $R$ is the identity matrix. The weight matrix $S$ has parameters $\alpha_1=10^3, \alpha_2=10^6$.  $U$ is initialized with $u_t \sim U[10^{-3},10^{-2}]$ and is fixed across experiments. \edits{We initialize $\eta_0=0.01$ from a small hyperparameter grid search.}
	
	\edits{We compare with A-optimality and L-optimality}:
		\begin{align}\label{eq:A_exp_des}
					\min_{U\in\mathcal{U}}\E_{X, \theta}\big[ (\hat{\theta}-\theta)^T H (\hat{\theta}-\theta)\big], 
			\end{align}
	 where $\hat{\theta}$ is a function of $X,\theta$ as in \eqref{eq:estimator} and $H$ is a positive semi-definite weight matrix that is the identity matrix in A-optimal design. For the L-optimal design, we use $H$ inspired from \cite{wagenmaker_task-optimal_2021} which considers the parameter sensitivity of the optimality gap  $\mathcal{R}(\pi;\theta) := J(\pi; \theta) - \inf_\pi J(\pi;\theta)$ under a policy $\pi$ such that $H= \nabla_\theta^2\mathcal{R}(\pi;\theta)\rvert_{\theta=\hat{\theta}} + \mu I$, with $\mu$ chosen to ensure positive semi-definiteness. We solve this using a gradient estimator of the same form as \eqref{eq:pathwise_mc}.
	
	\subsubsection{Results and Discussion}
\begin{figure}[th]
		\begin{center}
			\includegraphics[width=8.4cm]{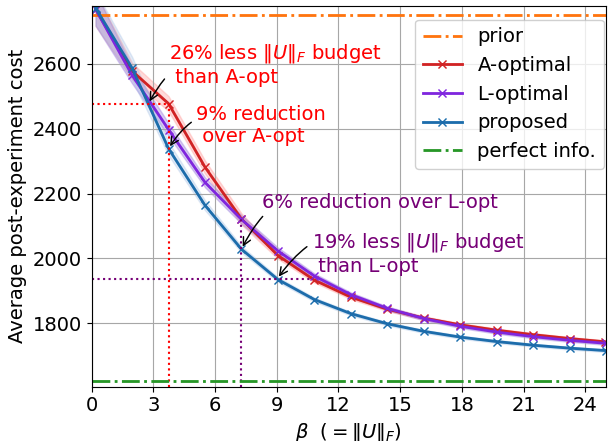}    % The printed column width is 8.4 cm.
			\caption{We compare the performance of our control-oriented system identification against A-optimal experiment design for a system with five states and three inputs, and known initial condition $x_0=[0., -4.3, 0., 2.1, 2.5]^T$ as in \cite{levine_optimal_1966}. \edits{The value of $\beta$ is varied and this constraint is active in all cases.} We include 95$\%$ confidence intervals using $10^5$ samples.} 
			\label{fig:norm_sweep}
		\end{center}
	\end{figure}

%	\edits{For control-oriented identification, determining performance on the ``true" system is often of interest and asymptotic convergence of parameter estimates to the ``true" parameters is desired. Alternative measures of performance include that on the ``worst-case" system for priors with finite support, and the expected value over all systems in the prior. In our work, we focused on the latter characterization as it accounts for performance over all systems that might be the ``true" system. However, we go on to make comparisons against robust methods and a ``true" system for comparison purposes with existing work.}
	We compare the performance of our method against A- and L- optimal design \eqref{eq:A_exp_des} in terms of post-experiment LQR control performance \eqref{eq:lqr}. For the experiment design \eqref{eq:exp_des}, we consider a feasible input set $\mathcal{U} = \left\{ U \,|\, \|U\|_F \leq \beta \right\}
	$, with $\beta$ a design parameter. We vary the allowed magnitude, $\beta$, in Figure \ref{fig:norm_sweep} and observe our method outperforms the alternative designs uniformly. \edits{More notably, the input budget needed to achieve the same cost as our method is significantly more for most values of $\beta$}. For any experiment design, if we knew the values of $(A,B)$, we would achieve the lowest possible control cost such that this is a lower bound on achievable performance. We also show the expected control performance associated with using a controller that uses the \emph{a-priori} system estimate, which indicates the gap for improvement via experimentation.
%	 \edits{\sout{instead of the \emph{a-posteriori} as in \eqref{eq:gen_controller}. Intuitively, as the budget $\beta$ for an experiment increases, so should the experiment performance as the experiment can ``explore more". For small $\beta$ the methods are nearly comparable as the additional information is minimal, but as $\beta$ increases the performance of our method approaches that of the perfect knowledge case. The A-optimal design only slowly decreases even though in the limit it should reach the lower bound.}}
	
\begin{figure}[th]
		\begin{center}
			\includegraphics[width=8.4cm]{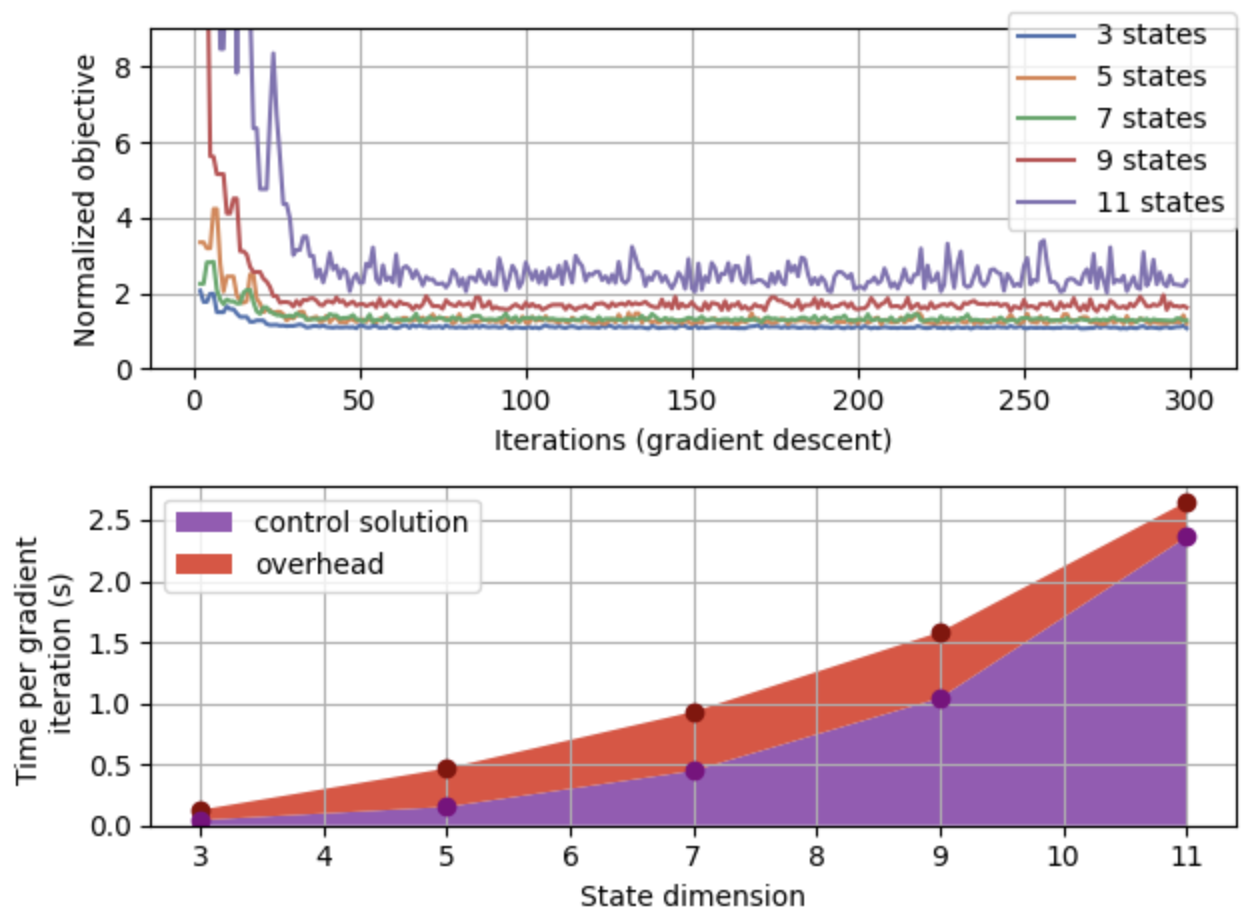}    % The printed column width is 8.4 cm.
			\caption{In the upper subplot, the number of iterations to converge for the car string problem is essentially the same regardless of system size suggesting good scaling properties of our method. In the lower subplot, we observe the average time to compute a single gradient sample \edits{on an Apple M1 Pro 10 core CPU with 32GB RAM in JAX \cite{jax2018github}}. The time is dominated by solving the control problem and ``overhead" refers to tasks such as automatic differentiation, initial compile time, etc.} 
			\label{fig:scaling}
		\end{center}
	\end{figure}

%	\edits{\sout{In Figure \ref{fig:input_comp}, we observe the optimal experiment inputs ($\Delta u$) for a three-car system. The experiment inputs for our control-oriented approach exhibit more excitation over the time horizon than the A-optimal design, which has a relatively smooth experiment input sequence, suggesting our method would perform better, which is verified by Figure \ref{fig:norm_sweep}. We also consider what the input sequence would look like if the controller given the \emph{a-priori} system estimate is used during an experiment trial. Since the controller is closed-loop \eqref{eq:ricc_diff}, we show the average input sequence. Using the optimal control may seem like a natural way to conduct an experiment, but the norm bound $\beta$ on the input is not active in this case, indicating that simply performing the optimal control leaves experiment budget unused.}}
%	
	Figure \ref{fig:scaling} shows how the problem scales with the system dimension. In the first subplot we see the convergence of the experiment criteria in \eqref{eq:exp_des} as a function of iterations. The criteria is normalized by the lower bound (given by the performance if we knew $A,B$). The number of iterations until the criteria stabilizes is roughly constant across problem dimension suggesting that the number of iterations required is independent of the system size \edits{though the variance tends to grow with system dimension. Since the convergence rate of SGD is closely tied to the Lipschitz constant, this would suggest that the Lipschitz constant is roughly the same as this car string problem scales.} In the second subplot, the time to compute each gradient sample is shown as a function of the state dimension. \edits{A- and L-optimal design avoid computing the post-experiment optimal control and control cost, such that the computation time should roughly be the red band in Figure \ref{fig:scaling}. However, the offline experiment design setting reduces the necessity of fast computation time.}
	
%	\edits{\sout{The compute time is dominated by ``control solution"--the time to solve the LQR problem--thus suggesting that the time-complexity is dominated by the control problem. ``Overhead" refers to the remainder of the compute tasks such as automatic differentiation and initialization of objects in the python library JAX \cite{noauthor_jax_2023}.}}
	
%	\begin{figure}[th]
%		\begin{center}
%			\includegraphics[width=7.4cm]{media/input_comp.png}    % The printed column width is 8.4 cm.
%			\caption{The experiment input sequences for each car in a 3 car system are compared under the proposed method, A-optimal experiment design, and by conducting the experiment with a feedback controller given the prior.} 
%			\label{fig:input_comp}
%		\end{center}
%	\end{figure}

	\subsection{\edits{Dual control as experiment design}}

	We also compare our method against a dual control approach \cite{umenberger2019} termed “robust reinforcement learning” (RRL) that approximately solves a problem of the form: 
		\begin{align}
			\min_{\pi_i^{RRL}} \sum_{i=1}^{N_{epochs}} \sup_{\theta \in \mathbf{\Theta}_i} \mathbb{E}_{w_t, e_t \forall t}\bigg[ \sum_{t=t_{i-1}}^{t_i }x_t^TQx_t + u_t^TRu_t\bigg]
		\end{align}
	where the dynamics \eqref{eq:linear_dynamics} are driven by $u_t^{RRL} = \pi^{RRL}(x_t) = Kx_t + \Sigma^{\frac{1}{2}}e_t, e_t \sim \mathcal{N}(0,I)$, with optimization variables ($K, \Sigma)$. The set $\mathbf{\Theta}_i$ contains system parameters such that $P(\theta_{tr} \in \mathbf{\Theta}_i \,|\, \mathcal{D}_i) = 1-\delta$. The dataset is initialized with  $\mathcal{D}_1=\mathcal{D}_{prior}$, an initial dataset gathered from $N_{traj}$ trajectories of a system $\theta_{tr}$. In this setup our prior is Gaussian with mean $\mathbb{E}_\theta[\theta \,|\, \mathcal{D}_{prior}]$ and variance $Var(\theta \,|\, \mathcal{D}_{prior})$ obtained from least squares estimation in RRL. $\pi^{RRL}$ can be considered an alternative experiment input signal to ours and we consider the application of $\pi^{RRL}$ and our method for a single epoch. We use the code and the problem setup from \cite{umenberger2019} with $\theta_{tr}$ given by:
	\begin{align} \label{eq:theta_true}
		A = \begin{bmatrix}
			1.1 & 0.5 & 0 \\
			0 & 0.9 & 0.1 \\
			0 & -0.2 & 0.8
		\end{bmatrix}, \quad
		B = \begin{bmatrix}
			0 & 1 \\
			0.1 & 0 \\
			0 & 2
		\end{bmatrix}
	\end{align}
	and $Q = I, R = \text{blkdiag}(0.1, 1), \sigma_w = 0.5, \delta=0.05$. The only parameters we change are the control and experiment horizon $T=N=20$, which improves the computation of sample averages for both approaches.
	
	Because our design requires a bound, $\beta$, on the experiment input and \cite{umenberger2019} does not include one, we first run the RRL experiments and then bound our design such that $\beta = \frac{1}{S}\sum_{s=1}^S \norm{U_{(s)}^{RRL}}_F$ similar to \cite{lee2023fundamental, wagenmaker_task-optimal_2021} where $U_{(s)}^{RRL}$ is an input sequence realization under RRL.
	
\begin{figure}[th]
			\begin{center}
					\includegraphics[width=8.4cm]{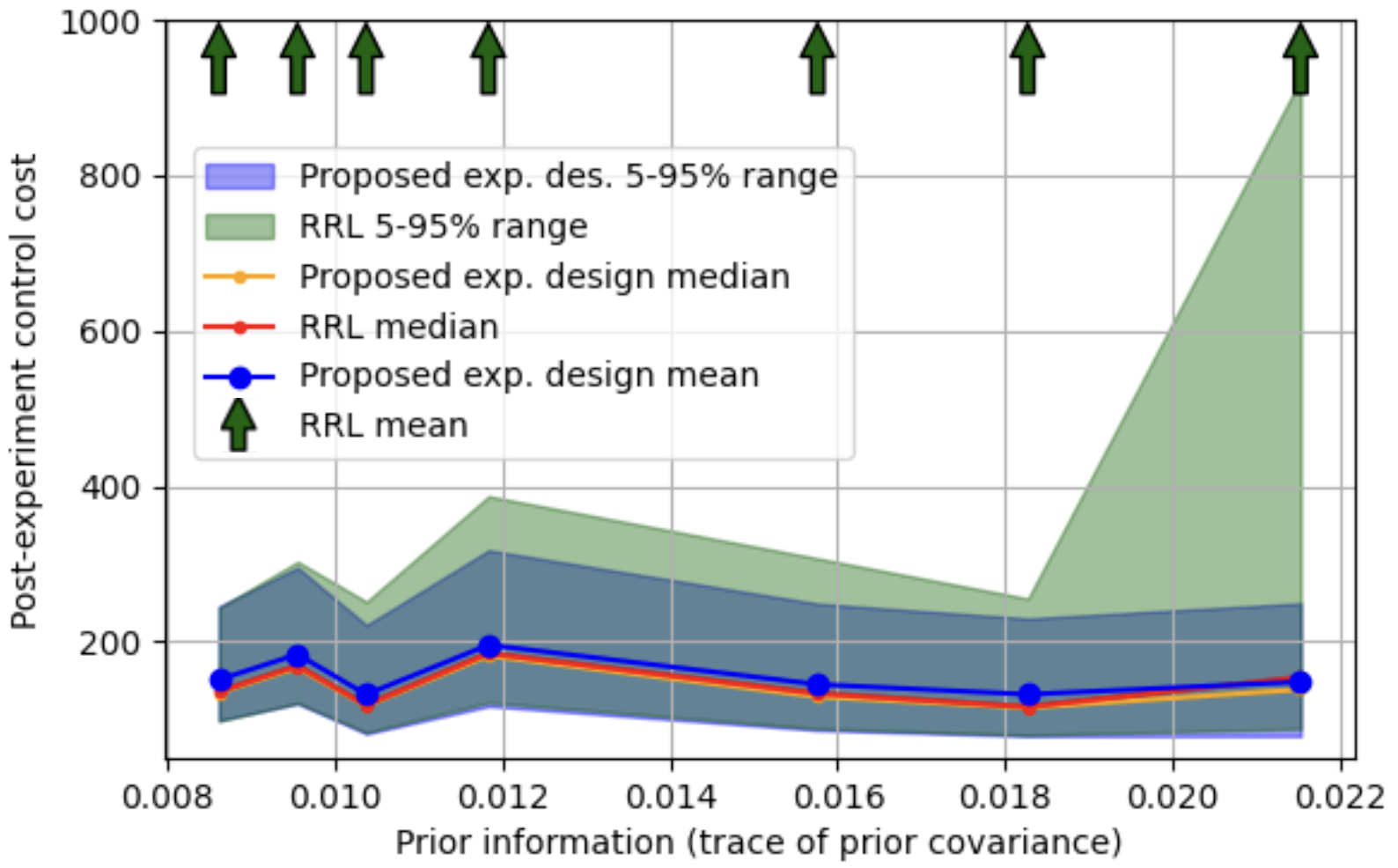}    % The printed column width is 8.4 cm.
					\caption{We show the performance of our proposed method against RRL by varying the prior information, which is achieved by varying $N_{traj}$ from 500 trajectories used in the \cite{umenberger2019} down to 200 in increments of 50. The mean value for our method is shown in blue where the post-experiment cost remains below 200 for all priors. For RRL, we observe that a few very large samples move the RRL mean to be very large such that we use arrows to indicate the values lie outside the axis limits.} 
					\label{fig:rrl_comp}
				\end{center}
	\end{figure}

	In Figure \ref{fig:rrl_comp} we vary the information in the prior--measured as the trace of the prior covariance--by varying $N_{traj}$ from 500 down to 200. RRL suffers from a very small percentage of systems causing the average to be very large and we also see the range of the realized costs from the 5th and 95th percentiles is smaller for our method. Finally, since a particular system \eqref{eq:theta_true} generates the dataset for RRL, we note that across the datasets our method improves on RRL by an average of 1\% and up to 4\%. In terms of computation time, RRL takes 1.5 seconds and ours 30.0 seconds, so RRL is more than an order of magnitude faster.
	
	%%%%%%%%%%%%%%%%%%%%%%%%%%%%%%%%%%%%%%%%%%%%%%%%%%%%%%%%%%%%%%%%%%%%%%%%%%%%%%%%
	\section{Conclusion}
	
	We proposed a control-oriented identification approach that in expectation improves data-driven controllers by construction. Our solution method via SGD is numerically shown in the LQR setting to outperform relevant benchmarks. Establishing results on the convergence rate and sample complexity of the stochastic gradient descent is important future work.

	\bibliographystyle{IEEEtran}
	\bibliography{IEEEabrv,bibliography}

\begin{thebibliography}{10}
\providecommand{\url}[1]{#1}
\csname url@samestyle\endcsname
\providecommand{\newblock}{\relax}
\providecommand{\bibinfo}[2]{#2}
\providecommand{\BIBentrySTDinterwordspacing}{\spaceskip=0pt\relax}
\providecommand{\BIBentryALTinterwordstretchfactor}{4}
\providecommand{\BIBentryALTinterwordspacing}{\spaceskip=\fontdimen2\font plus
\BIBentryALTinterwordstretchfactor\fontdimen3\font minus
  \fontdimen4\font\relax}
\providecommand{\BIBforeignlanguage}[2]{{%
\expandafter\ifx\csname l@#1\endcsname\relax
\typeout{** WARNING: IEEEtran.bst: No hyphenation pattern has been}%
\typeout{** loaded for the language `#1'. Using the pattern for}%
\typeout{** the default language instead.}%
\else
\language=\csname l@#1\endcsname
\fi
#2}}
\providecommand{\BIBdecl}{\relax}
\BIBdecl

\bibitem{chernoff_locally_1953}
H.~Chernoff, ``Locally {Optimal} {Designs} for {Estimating} {Parameters},''
  \emph{The Annals of Mathematical Statistics}, vol.~24, no.~4, pp. 586--602,
  Dec. 1953, publisher: Institute of Mathematical Statistics.

\bibitem{simon_dynamic_1956}
H.~A. Simon, ``Dynamic {Programming} {Under} {Uncertainty} with a {Quadratic}
  {Criterion} {Function},'' \emph{Econometrica}, vol.~24, no.~1, pp. 74--81,
  1956, publisher: [Wiley, Econometric Society].

\bibitem{elfving_optimum_1952}
G.~Elfving, ``Optimum {Allocation} in {Linear} {Regression} {Theory},''
  \emph{The Annals of Mathematical Statistics}, vol.~23, no.~2, pp. 255--262,
  Jun. 1952, publisher: Institute of Mathematical Statistics.

\bibitem{lindqvist_identification_2001}
K.~Lindqvist and H.~Hjalmarsson, ``\BIBforeignlanguage{en}{Identification for
  control: adaptive input design using convex optimization},'' in
  \emph{\BIBforeignlanguage{en}{Proceedings of the 40th {IEEE} {Conference} on
  {Decision} and {Control} ({Cat}. {No}.{01CH37228})}}, vol.~5.\hskip 1em plus
  0.5em minus 0.4em\relax Orlando, FL, USA: IEEE, 2001, pp. 4326--4331.

\bibitem{gevers_identification_2005}
M.~Gevers, ``Identification for {Control}: {From} the {Early} {Achievements} to
  the {Revival} of {Experiment} {Design},'' in \emph{Proceedings of the 44th
  {IEEE} {Conference} on {Decision} and {Control}}, Dec. 2005, pp. 12--12.

\bibitem{anderson_cdc2023}
S.~Anderson, K.~Byl, and J.~P. Hespanha, ``Experiment design with {Gaussian}
  process regression with applications to chance-constrained control,'' in
  \emph{2023 62nd IEEE Conference on Decision and Control (CDC)}, 2023, pp.
  3931--3938.

\bibitem{lee2023fundamental}
B.~D. Lee, I.~Ziemann, A.~Tsiamis, H.~Sandberg, and N.~Matni, ``The fundamental
  limitations of learning linear-quadratic regulators,'' in \emph{2023 62nd
  IEEE Conference on Decision and Control (CDC)}.\hskip 1em plus 0.5em minus
  0.4em\relax IEEE, 2023, pp. 4053--4060.

\bibitem{dean_sample_2020}
S.~Dean, H.~Mania, N.~Matni, B.~Recht, and S.~Tu, ``\BIBforeignlanguage{en}{On
  the {Sample} {Complexity} of the {Linear} {Quadratic} {Regulator}},''
  \emph{\BIBforeignlanguage{en}{Foundations of Computational Mathematics}},
  vol.~20, no.~4, pp. 633--679, Aug. 2020.

\bibitem{simchowitz_naive_2020}
M.~Simchowitz and D.~Foster, ``\BIBforeignlanguage{en}{Naive {Exploration} is
  {Optimal} for {Online} {LQR}},'' in \emph{\BIBforeignlanguage{en}{Proceedings
  of the 37th {International} {Conference} on {Machine} {Learning}}}.\hskip 1em
  plus 0.5em minus 0.4em\relax PMLR, Nov. 2020, pp. 8937--8948, iSSN:
  2640-3498.

\bibitem{astrom_adaptive_2008}
K.~J. Åström and B.~Wittenmark, \emph{Adaptive control}.\hskip 1em plus 0.5em
  minus 0.4em\relax Courier Corporation, 2008.

\bibitem{umenberger2019}
J.~Umenberger, M.~Ferizbegovic, T.~B. Sch\"{o}n, and H.~k. Hjalmarsson,
  ``Robust exploration in linear quadratic reinforcement learning,'' in
  \emph{Advances in Neural Information Processing Systems}, vol.~32.\hskip 1em
  plus 0.5em minus 0.4em\relax Curran Associates, Inc., 2019.

\bibitem{ferizbegovic2020}
M.~Ferizbegovic, J.~Umenberger, H.~Hjalmarsson, and T.~B. Schön, ``Learning
  robust {LQ}-controllers using application oriented exploration,'' \emph{IEEE
  Control Systems Letters}, vol.~4, no.~1, pp. 19--24, 2020.

\bibitem{venkatasubramanian2020}
J.~Venkatasubramanian, J.~Köhler, J.~Berberich, and F.~Allgöwer, ``Robust
  dual control based on gain scheduling,'' in \emph{2020 59th IEEE Conference
  on Decision and Control (CDC)}, 2020, pp. 2270--2277.

\bibitem{rallo2018}
G.~Rallo, S.~Formentin, C.~R. Rojas, and S.~M. Savaresi, ``Robust experiment
  design for virtual reference feedback tuning,'' in \emph{2018 IEEE Conference
  on Decision and Control (CDC)}, 2018, pp. 2271--2276.

\bibitem{mohamed_monte_2020}
S.~Mohamed, M.~Rosca, M.~Figurnov, and A.~Mnih, ``Monte {Carlo} {Gradient}
  {Estimation} in {Machine} {Learning},'' Sep. 2020.

\bibitem{peters_reinforcement_2008}
J.~Peters and S.~Schaal, ``\BIBforeignlanguage{en}{Reinforcement learning of
  motor skills with policy gradients},'' \emph{\BIBforeignlanguage{en}{Neural
  Networks}}, vol.~21, no.~4, pp. 682--697, 2008.

\bibitem{fan_fast_2017}
K.~Fan, Z.~Wang, J.~Beck, J.~Kwok, and K.~A. Heller, ``Fast second order
  stochastic backpropagation for variational inference,'' in \emph{Advances in
  Neural Information Processing Systems}, vol.~28.\hskip 1em plus 0.5em minus
  0.4em\relax Curran Associates, Inc., 2015.

\bibitem{rossi2006}
P.~E. Rossi, G.~M. Allenby, and R.~McCulloch, \emph{Bayesian Statistics and
  Marketing}.\hskip 1em plus 0.5em minus 0.4em\relax Wiley, Oct. 2006, pp.
  31--34.

\bibitem{bertsekas_dynamic_2012}
D.~Bertsekas, \emph{Dynamic programming and optimal control: {Volume}
  {I}}.\hskip 1em plus 0.5em minus 0.4em\relax Athena scientific, 2012, vol.~4,
  pp. 110--112.

\bibitem{levine_optimal_1966}
W.~Levine and M.~Athans, ``On the optimal error regulation of a string of
  moving vehicles,'' \emph{IEEE Transactions on Automatic Control}, vol.~11,
  no.~3, pp. 355--361, Jul. 1966.

\bibitem{anderson_tech24}
\BIBentryALTinterwordspacing
S.~Anderson and J.~P. Hespanha, ``Control-oriented identification for the
  linear quadratic regulator: Technical report,'' Santa Barbara, Mar. 2024.
  [Online]. Available: \url{https://arxiv.org/abs/2403.05455}
\BIBentrySTDinterwordspacing

\bibitem{wagenmaker_task-optimal_2021}
A.~J. Wagenmaker, M.~Simchowitz, and K.~Jamieson,
  ``\BIBforeignlanguage{en}{Task-{Optimal} {Exploration} in {Linear}
  {Dynamical} {Systems}},'' in \emph{\BIBforeignlanguage{en}{Proceedings of the
  38th {International} {Conference} on {Machine} {Learning}}}.\hskip 1em plus
  0.5em minus 0.4em\relax PMLR, Jul. 2021, pp. 10\,641--10\,652, iSSN:
  2640-3498.

\bibitem{jax2018github}
J.~Bradbury, R.~Frostig, P.~Hawkins, M.~J. Johnson, C.~Leary, D.~Maclaurin,
  G.~Necula, A.~Paszke, J.~Vander{P}las, S.~Wanderman-{M}ilne, and Q.~Zhang,
  ``{JAX}: composable transformations of {P}ython+{N}um{P}y programs,'' 2018.

\end{thebibliography}

	%	\section{ACKNOWLEDGMENTS}
	%	
	%	The authors gratefully acknowledge the National Science Foundation Graduate Research Fellowship.
	
	\section{Appendix}
	
	\subsection{Car String Setting} \label{sec:car_string_app}
	
	For the car string problem with three cars, $A$ and $B$ are given by:
	\begin{subequations}
	\begin{align}
		A &= \begin{bmatrix}
			-\frac{\alpha^{(1)} T_s}{m^{(1)}} + 1 & 0 & 0 & 0 &\hdots \\
			T_s & 1 & -T_s & 0 & \hdots  \\
			0&0&-\frac{\alpha^{(2)}T_s}{m^{(2)}} +1& 0 & \hdots \\
			0 & 0 & T_s & 1 &\hdots \\
			\vdots & \vdots & \vdots & \vdots& \ddots
		\end{bmatrix} \\
		% Matrix Bv
		B &=  \begin{bmatrix}
			\frac{T_s}{m^{(1)}} & 0 & \hdots \\
			0 & 0  & \hdots \\
			0&\frac{T_s}{m^{(2)}} & \hdots \\
			0 & 0  & \hdots\\
			\vdots&\vdots & \ddots
		\end{bmatrix}.
	\end{align}
	\end{subequations}
	
	The prior on the parameters for \eqref{eq:estimator} is $\Theta_0=[A,B]$ with $m^{(1)}=m^{(2)}=m^{(3)}=1$, $\alpha^{(1)}=\alpha^{(2)}=\alpha^{(3)}=1$, $T_s=0.1$; the prior covariance is set to $\Lambda_0^{-1}=\text{diag}([0.1, 0.01, 0.05, 0.1, 0.01, 0.05, 0.1, 0.05])$.
	
	\subsection{Change of variable for linear system} \label{sec:linear_change_of_var}
	We want to find a change of variable for the linear system \eqref{eq:linear_dynamics}. We start by showing the case for $t=2$ such that
	\begin{subequations}
		\begin{align}
			x_{1} &= Ax_0 + Bu_0 + w_0 \\
			x_{2} &= Ax_1 + Bu_1 + w_1 \\
			\intertext{and expressing $x_1$ in terms of $x_0$}
			&=A^2x_0 + ABu_0 + Aw_0+ Bu_1 + w_1
		\end{align}
		Then assuming this holds for time $t$:
		\begin{align}  \label{eq:rec_t}
			x_t = A^tx_0 + \sum_{l=0}^{t-1}A^{t-1-l}(Bu_l + w_l),
		\end{align}
		at time $t+1$
		\begin{align} 
			x_{t+1} &= Ax_t + Bu_t + w_t \\
			\intertext{substituting the recursion \eqref{eq:rec_t}}
			x_{t+1} &= A( A^tx_0 + \sum_{l=0}^{t-1}A^{t-1-l}(Bu_l + w_l)) + B_t + w_t, \\
			x_{t+1} &= A^{t+1}x_0 + \sum_{l=0}^{t}A^{t-l}(Bu_l + w_l),
		\end{align}
	\end{subequations}
	the desired result, where $w_t$ is distributed according to the process noise and is independent of $u_t$.

	\subsection{Differentiability of the value function} \label{sec:diff_app}
	
	We specialize the form of the gradient in \eqref{eq:pw_mc} to the LQR setting (Section \ref{sec:oed_lqr}) and show the differentiability assumptions are met. In particular, with the certainty equivalent control $\pi:=\{K_0,...,K_{N-1}\}$ as in \eqref{eq:ricc_diff}, constructed from the estimate \eqref{eq:estimator}, we parameterize the controller in terms of the certainty equivalent estimate such that $\pdv{J}{\pi_s} = \pdv{J}{K_{tqr}}\pdv{K_{tqr}}{\hat{\theta}}$ and $(\pdv{K_{s}}{X_{mn}}\pdv{g_{mn}}{U_{ij}} + \pdv{K_{s}}{U_{ij}})=\pdv{\hat{\theta}_{mn}}{u_{ij}}$. In the LQR setting this parameterization is convenient based on the forms of the objective,  controller, and estimator as it differentiates \eqref{eq:lqr}, \eqref{eq:ricc_diff}, and \eqref{eq:estimator}. This leads to the $ij$th element of the gradient for a single sample as
	\begin{align}
		\pdv{J}{u_{ij}} = \sum_{t,q,r,m,n} \pdv{J}{K_{tqr}} \pdv{K_{tqr}}{\hat{{\Theta}}_{mn}} \pdv{\hat{\Theta}_{mn}}{u_{ij}}
	\end{align}
	where the summation is over all the dimensions of the feedback gains and estimator. In the following, we address each component of the gradient separately.
	\subsubsection{Gradient of $J$ with respect to $K_t$}
	First, we observe that the gradient of \eqref{eq:closed_form_cost} as derived in the subsequent section at $t=0$ is given by 
	\begin{subequations}
		\begin{align}
			\pdv{J}{K_0} &= 2\big((R+B^TP_1B)K_0 + B^TP_1A\big)x_0x_0^T
		\end{align}
		and for $t>0$ as
		\begin{align}
			\begin{split}
				&\pdv{J}{K_t} = 2\big[(R+ B^TP_{t+1}B)K_{t} + B^TP_{t+1}A\big] \\
				&\times  \bigg(\Pi_{i=0}^{t-1}(A+BK_i)x_0x_0^T\Pi_{i=t-1}^{0}(A+BK_i)^T + \Sigma_w \\
				&+ \sum_{j=1, t>1}^{t-1} \Pi_{i=j}^{t-1}(A+BK_i)\Sigma_w\Pi_{i=t-1}^{j}(A+BK_i)^T \bigg).
			\end{split}
		\end{align}
	\end{subequations}
	For a finite horizon, the entries of $P_t$ are finite even if the cost grows exponentially in time such that the gradient itself will be finite. If we want to bound the gradient, the gradient is polynomial in the Gaussian random variables $(A,B)$ such that there exists a polynomial function of the random variables, which is integrable.
	\subsubsection{Derivation: Gradient of $J$ with respect to $K_t$}
	We want to take the gradient of the value function 
	\begin{subequations}
		\begin{align}
			J(\Theta, \pi) := x_0^T P_0 x_0 + \sum_{t=0}^{N-1}\tr(P_{t+1}\Sigma_W)
		\end{align}
		with respect to $K_t$. For $K_0$ we expand $P_0$ to see the dependence
		\begin{align}
			\begin{split}
				&\pdv{J(\Theta, \pi)}{K_0} = \pdv{}{K_0} \bigg(x_0^T(Q + K_0^TRK_0 + \\ &(A+BK_0)^TP_1(A+BK_0)x_0 + \sum_{t=0}^{N-1}\tr(P_{t+1}\Sigma_W)\bigg).
			\end{split}
		\end{align}
		Evaluating this, we get
		\begin{align}
			\pdv{J(\Theta, \pi)}{K_0} = (2RK_0 + 2B^TP_1BK_0)x_0x_0^T + 2B^TP_1Ax_0x_0^T,
		\end{align}
		which can be rearranged to give the desired result.
		For $t>0$, there is dependence in both the initial condition and the process noise term. For the initial condition term, recursively expand $P_i$ until $i=t$, and then take the gradient as for $K_0$. If we define the current state as $x_t:=\Pi_{i=0}^{t-1}(A+BK_i)x_0$, then we can express this relationship as
		\begin{align}
			\pdv{(x_0^TP_0x_0)}{K_t} = 2\big[(R+ B^TP_{t+1}B)K_{t} + B^TP_{t+1}A\big]x_tx_t^T.
		\end{align}
		This gives us the first part of the gradient. The second part is due to the process noise and follows a similar pattern. Start by expanding $P_{t+1}$ to get terms of $K_{t+1}$:
		\begin{align}
			\begin{split}
				&\trace(P_{t}\Sigma_w) = \\
				&\trace\big((Q+ K_t^TRK_t + (A+BK_t)^TP_{t+1}(A+BK_t))\Sigma_w\big)
			\end{split}
		\end{align}
		%and taking the gradient of each term since $\trace(A+B) = \trace(A) + \trace(B)$, gives us
		%\begin{align}
		%	\pdv{}{K_{t+1}} \big(\trace(K_t\Sigma_wK_t^TR)\big) = 0
		%\end{align}
		%\begin{align}
		%	&\pdv{}{K_{t+1}} \big(K_t \Sigma_w K_t^T B^TP_{t+1}B\big) =  \pdv{}{K_{t+1}} \big(\alpha P_{t+1}\big)
		%\end{align} 
		%using the cyclic property and  $\alpha:=B K_t \Sigma_w K^T B^T$. Similarly,
		%\begin{align}
		%	\pdv{}{K_{t+1}} 2\trace(\Sigma_w A^TP_{t+1}BK_t) = \pdv{}{K_{t+1}} 2\trace(\alpha_2 P_{t+1})
		%\end{align}
		%with $\alpha_2 = BK_t\Sigma_w A^T$ and
		%\begin{align}
		%	\pdv{}{K_{t+1}} \trace(\Sigma_w A^TP_{t+1}A) = \pdv{}{K_{t+1}} 2\trace(\alpha_3 P_{t+1})
		%\end{align}
		%with $\alpha_3:= A \Sigma_w A^T$.
		Expanding $P_{t+1}$, we need to take gradients of the following terms (here given at $t$):
		\begin{align}
			&\pdv{K_t} \trace(K_t \Sigma K_t^TR) = 2RK_t\Sigma_w.\\
			&\pdv{K_t} \trace(K_t \Sigma_w K_t^T B^TP_{t+1}B) = 2B^TP_{t+1}BK_t \Sigma_w. \\
			&\pdv{K_t} 2\trace(\Sigma_w A^TP_{t+1}BK_t) = 2B^TP_{t+1}A\Sigma_w.
		\end{align}
	\end{subequations}
	Using these gradients and algebraic manipulations, we get the desired result for one step for the process noise term. This can be repeated for all time steps to get the overall result. Combining the initial condition terms with the noise terms gives us the gradient for $t>0$.
	%Applying this to the terms of interest with $\bar{\alpha}=(\alpha_1+\alpha_2)$, we get
	%\begin{align}
	%	\begin{split}
		%	&\pdv{}{K_{t+1}} \big( \bar{\alpha}P_{t+1}\big) = 2\bigg(RK_{t+1}\bar{\alpha} + B^TP_{t+2}BK_{t+1}\bar{\alpha} \\
		%	&+  B^TP_{t+2}A\Sigma_w\bar{\alpha}\bigg) \\
		%	& = 2\big((R+B^TP_{t+2}B)K_{t+1} + B^TP_{t+2}A\big)\Sigma_w\bar{\alpha} 
		%\end{split} \\
		%\end{align}
		%FINISH
		
		\subsubsection{Gradient of $K_t$ with respect to estimate $\Theta$}
		Next, we examine the gradient of the data-driven control with respect to the estimated system as derived in the next section, denoted above as $(\hat{A},\hat{B})$ as we use certainty equivalence in the dynamics parameters. The gradient of 
		\begin{subequations}
			\begin{align}
				K_t = -(R+\hat{B}^TP_{t+1}\hat{B})^{-1}\hat{B}^TP_{t+1}\hat{A}
			\end{align}
			with respect to $\hat{\Theta}$ is most easily written in terms of the elements of $\hat{A},\hat{B}$.
			
			The gradient is recursively computed as 
			\begin{align}
				\begin{split}
					\pdv{K_t}{A_{ij}} &= (R + B^TP_{t+1}B)^{-1}B^T\pdv{P_{t+1}}{A_{ij}}B(R + B^TP_{t+1}B)^{-1} 
					\\ &- (R + B^TP_{t+1}B)^{-1}\bigg(B^T\pdv{P_{t+1}}{A_{ij}}A+B^TP_{t+1}e_{ij}\bigg)
				\end{split} \\
				\begin{split}
					\pdv{K_t}{B_{ij}} &= (R + B^TP_{t+1}B)^{-1}\bigg(2e_{ij}^TP_{t+1}B + B^T\pdv{P_{t+1}}{B_{ij}}B\bigg)\\
					&\times (R + B^TP_{t+1}B)^{-1} \\ &- (R + B^TP_{t+1}B)^{-1}\bigg(B^T\pdv{P_{t+1}}{B_{ij}}B+2e_{ij}^TP_{t+1}B\bigg)
				\end{split}
			\end{align}
			with
			\begin{align}
				\begin{split}
					\pdv{P_t}{A_{ij}} &= 2 \pdv{K_t}{A_{ij}}^TRK_t + A^T\pdv{P_{t+1}}{A_{ij}}A \\
					& + 2e_{ij}^TP_{t+1}A + 2e_{ij}^TP_{t+1}BK_t \\
					&+ 2A^T\pdv{P_{t+1}}{A_{ij}}BK_t + 2A^TP_{t+1}B\pdv{K_t}{A_{ij}} \\
					&+2\pdv{K_t}{A_{ij}}^TB^TP_{t+1}BK_t + K_t^TB^T\pdv{P_{t+1}}{A_{ij}}BK_t
				\end{split} \\
				\begin{split}
					\pdv{P_t}{B_{ij}} &= 2 \pdv{K_t}{B_{ij}}^TRK_t + A^T\pdv{P_{t+1}}{B_{ij}}A +   2A^TP_{t+1}e_{ij}K_t \\
					&+ 2A^T\pdv{P_{t+1}}{B_{ij}}BK_t 
					+ 2A^TP_{t+1}B\pdv{K_t}{B_{ij}} \\
					&+2\pdv{K_t}{B_{ij}}^TB^TP_{t+1}BK_t + K_t^TB^T\pdv{P_{t+1}}{B_{ij}}BK_t \\
					&+ 2K_t^Te_{ij}^TP_{t+1}BK_t.
				\end{split}
			\end{align}
		\end{subequations}
		with $P_N=Q_N$. If we want to bound the gradient, the elements of $P_t$ as governed by \eqref{eq:ricc_diff} will be finite for a finite horizon, leading to finite values for the gradients. Furthermore, the gradient is again polynomial in the parameters such that there exists a polynomial function that upper bounds the gradient and is integrable.
		
		\subsubsection{Derivation: Gradient of $K_t$ with respect to estimate $\Theta$}
		For a posterior distribution with mean $\hat{\Theta}=[\hat{A},\hat{B}]$, and the controller defined by the Riccati difference equations:
		\begin{subequations}
			\begin{align} 
				K_t &= -(R+\hat{B}^TP_{t+1}\hat{B})^{-1}\hat{B}^TP_{t+1}\hat{A}, \\
				P_t &= Q + K_t^TRK_t - (\hat{A}+\hat{B}K_t)^TP_{t+1}(\hat{A}+\hat{B}K_t), 
			\end{align}
		\end{subequations}
		we want to find the gradient with respect to elements of $\hat{A}$ and $\hat{B}$. Starting with $K_t$:
		\begin{subequations}
			\begin{align}
				\pdv{K_t}{\hat{A}_{ij}} &= \pdv{\hat{A}_{ij}}\bigg( -(R+\hat{B}^TP_{t+1}\hat{B})^{-1}\hat{B}^TP_{t+1}\hat{A}\bigg) \\
				\begin{split}
					&= \pdv{\hat{A}_{ij}}\big( -(R+\hat{B}^TP_{t+1}\hat{B})^{-1}\big)\hat{B}^TP_{t+1}\hat{A} \\
					&+ \big( -(R+\hat{B}^TP_{t+1}\hat{B})^{-1}\big)  \pdv{\hat{A}_{ij}}\big(\hat{B}^TP_{t+1}\hat{A}\big)
				\end{split}
			\end{align}
			For the gradient of the first component:
			\begin{align}
				&\pdv{\hat{A}_{ij}}\big( -(R+\hat{B}^TP_{t+1}\hat{B})^{-1}\big) = \\
				& (R+\hat{B}^TP_{t+1}\hat{B})^{-1}  \pdv{(R+\hat{B}^TP_{t+1}\hat{B})}{\hat{A}_{ij}} \big((R+\hat{B}^TP_{t+1}\hat{B})^{-1}\big) \\
				&= (R+\hat{B}^TP_{t+1}\hat{B})^{-1}  \hat{B}^T\pdv{P_{t+1}}{\hat{A}_{ij}}\hat{B} \big((R+\hat{B}^TP_{t+1}\hat{B})^{-1}\big)
			\end{align}
			and the second
			\begin{align}
				\pdv{\hat{A}_{ij}}\big(\hat{B}^TP_{t+1}\hat{A}\big) = 
				\hat{B}^T\bigg(\pdv{P_{t+1}}{\hat{A}_{ij}}\hat{A} + P_{t+1}e_{ij}\bigg).
			\end{align}
			The results for the partial with respect to $B_{ij}$ follows similarly. In each case, we need to compute the partial of $P_{t}$:
			\begin{align}
				\begin{split}
					&\pdv{P_t}{\hat{A}_{ij}} = \\
					&\pdv{\hat{A}_{ij}}\big(Q + K_t^TRK_t - (\hat{A}+\hat{B}K_t)^TP_{t+1}(\hat{A}+\hat{B}K_t)\big).
				\end{split}
			\end{align}
			$Q$ is independent of $\hat{A}$ (and $\hat{B}$). The rest of terms are:
			\begin{align}
				\pdv{\hat{A}_{ij}}\big(K_t^TRK_t\big) = 2\pdv{K_t}{\hat{A}_{ij}}^TRK_t,
			\end{align}
			\begin{align}
				&\pdv{\hat{A}_{ij}}\big((\hat{A}+\hat{B}K_t)^TP_{t+1}(\hat{A}+\hat{B}K_t)\big) = \\
				& 2e_{ij}^TP_{t+1}\hat{A} + \hat{A}^T\pdv{P_{t+1}}{\hat{A}_{ij}}\hat{A} + \pdv{\hat{A}_{ij}}\big(\hat{A}^TP_{t+1}\hat{B}K_t\big),
			\end{align}
			and
			\begin{align}
				\begin{split}
					&\pdv{\hat{A}_{ij}}\big(\hat{A}^TP_{t+1}\hat{B}K_t\big) = 2e_{ij}^TP_{t+1}\hat{B}K_t \\
					&+ 2\hat{A}^T\pdv{P_{t+1}}{\hat{A}_{ij}}\hat{B}K_t
					+2\hat{A}^TP_{t+1}\hat{B}\pdv{K_t}{\hat{A}_{ij}},
				\end{split}
			\end{align}
			A similar derivation follows with respect to $\hat{B}_{ij}$.
		\end{subequations}
		
		\subsubsection{Gradient of estimate $\Theta$ with respect to $U$ with derivation}
		Finally, to address the gradient of the estimated value with respect to the design variable $U \in \mathbb{R}^{n_u \times T}$ with entries $u_{ij}$, we first rewrite the estimator in \eqref{eq:estimator} using sums as
		\begin{subequations}
			\begin{align}
				\hat{\Theta} &= (\Theta_0 \Lambda_0+ \sum_{t=0}^{T-1} y_ts_tz_t^T) (\Lambda_0 + \sum_{t=0}^{T-1}z_ts_tz_t^T)^{-1}, \\
				&=: \Psi \Delta,
			\end{align}
			where
			\begin{align}
				y_t &= x_{t+1} = Ax_t + B\gamma_t, \\
				x_t &= A^{t}x_0 + \sum_{l=0}^{t-1}A^{t-1-l}(Bu_l + w_l), \\
				z_t &= [x_t; u_t].
			\end{align}
			
			$y_t,s_t,z_t$ all depend on $U$. We use the $\vect{\cdot}$ operator, which stacks the columns on tops of each other, to simplify the derivation. As such,
			\begin{align}
				\begin{split}
					\vect{\pdv{\hat{\Theta}}{u_{ij}}} &= (\Delta^T \otimes I) \vect{\pdv{\Psi}{u_{ij}}} + (I \otimes \Psi) \vect{\pdv{\Delta}{u_{ij}}},
				\end{split}
			\end{align}
			where
			\begin{align}
				\begin{split}
					&\vect{\pdv{\Psi}{u_{ij}}} = \\
					&\vect{\pdv{\sum_{t=0}^{T-1} y_ts_tz_t^T}{u_{ij}}} 
					= \sum_{t=0}^{T-1} \vect{\pdv{y_ts_tz_t^T}{u_{ij}}}, \\
					&= (z_ts_t^T \otimes I)\vect{\pdv{y_t}{u_{ij}}} + (z_t \otimes y_t)\vect{\pdv{s_t}{u_{ij}}}  \\
					&+ (I \otimes y_ts_t)\vect{\pdv{z_t^T}{u_{ij}}}.
				\end{split}
			\end{align}
			Each gradient in the above expression is
			\begin{align}
				\begin{split}
					\vect{\pdv{y_t}{u_{ij}}} &= \vect{\pdv{}{u_{ij}}A^{t+1}x_0 + \sum_{l=0}^{t}A^{t-l}(B\gamma_l + w_l)}, \\
					&= \sum_{l=0}^t(I \otimes A^{t-l}B)\vect{\pdv{\gamma_l}{u_{ij}}}, \\
					&= (I \otimes A^{t-i}B)\vect{e_{ij}}, \quad (t\geq i)
				\end{split} 
			\end{align}
			\begin{align}
				\begin{split}
					&\vect{\pdv{s_t}{u_{ij}}} =\vect{\pdv{}{u_{ij}}\atan((\norm{x_t}-\alpha_1)\alpha_2)/\pi + 0.5} \\
					&=\vect{\frac{1}{\pi \alpha_2^2} \frac{1}{1/\alpha_2^2 + (\norm{x_t}-\alpha_1)^2} \pdv{}{u_{ij}} \big((\norm{x_t}-\alpha_1)^2\big)} \\
					&\pdv{}{u_{ij}} \big((\norm{x_t}-\alpha_1)^2\big) = 2(\norm{x_t}-\alpha_1)\frac{x_t}{\norm{x_t}}\pdv{x_t}{u_{ij}}
				\end{split}
			\end{align}
			\begin{align}
				\begin{split}
					&\vect{\pdv{z_t^T}{u_{ij}}}= \\
					& \vect{\pdv{}{u_{ij}} \begin{bmatrix}
							A^{t}x_0 + \sum_{l=0}^{t-1}A^{t-1-l}(Bu_l + w_l) \\
							u_t
						\end{bmatrix}^T} \\
					&= \begin{bmatrix}
						(I \otimes A^{t-1-i}B)\vect{\pdv{u_i}{u_{ij}}}, \quad (t\geq i)\text{ else 0} \\
						\vect{e_{ij}}, \quad (i=t), \text{ else 0}
					\end{bmatrix}^T
				\end{split}
			\end{align}
			Going to the second term, $\Delta$,  in the estimator, we obtain
			\begin{align}
				&\vect{\pdv{\Delta}{u_{ij}}} = -(\Delta \otimes \Delta)\vect{\pdv{}{u_{ij}}\sum_{t=0}^{T-1}z_ts_tz_t^T}\\
				&= -(\Delta \otimes \Delta)\sum_{t=0}^{T-1} \Bigg((z_ts_t^T \otimes I) \\
				&\times \begin{bmatrix}
					(I \otimes A^{t-1-i}B)\vect{e_{ij}}, ~(t\geq j),\text{else 0}  \\
					\vect{e_{ij}}, ~(j=t), \text{else 0}
				\end{bmatrix} \\
				& +(z_t \otimes z_t)\vect{\pdv{s_t}{\gamma_i}} + (I\otimes Z_Ts_t) \\
				&\times  \begin{bmatrix}
					(I \otimes A^{t-1-i}B)\vect{e_{ij}}, ~(t\geq j),\text{else 0}  \\
					\vect{e_{ij}}, ~(i=t), \text{else 0}
				\end{bmatrix}^T\Bigg)
			\end{align}
			The gradient is then well-defined except if $\Delta$ were to be ill-defined due to lack of invertibility; however, the prior $\Lambda_0$ is chosen to be non-singular and obviates this possibility. The resulting expression contains a rational and polynomial term, such that there exists a polynomial bounding function. 
		\end{subequations}
		
		%%%%%%%%%%%%%%%%%%%%%%%%%%%%%%%%%%%%%%%%%%%%%%%%%%%%%%%%%%%%%%%%%%%%%%%%%%%%%%%

	\end{document}